\documentclass[prl,showpacs, twocolumn]{revtex4}
\usepackage[dvips]{graphicx}
\begin{document}
\title{Naturalness and Naturalness Criteria}

\author{Su Yan}
\email{yans@northwestern.edu}
\affiliation{Department of Physics and Astronomy,
Northwestern University, Evanston, IL 60201}
\date{\today}

\begin{abstract}
We analysis how to describe the level of naturalness and pointed out that  
Barbieri and Giudice's the widely adopted sensitivity criteria 
of naturalness can not reflect the level of naturalness correctly,
we analysis the problems of the sensitivity criteria and proposed 
a new criterion that can solve these problems, and also give a clear
physical meaning to the naturalness cut-off level.
\end{abstract}
 

\maketitle

The physical principle of naturalness introduced by Wilson and 
't Hooft\cite{Wilson} requires that in order to get a small observable parameters at
the weak scale,  
we do not need to extremely fine-tune the lagrangian parameters at the 
grand unification scale. 
For example, the renormalization of \(\Phi^4\) model:

\begin{equation}
\mathcal{L}=\frac{1}{2}[(\partial_{\mu}\Phi)^2-m^2_0\Phi^2]-\frac{g}{4!}\Phi^4
\label{equ-1}
\end{equation}

the scalar mass \(m^2\) can be written as:
\begin{equation}
m^2=m^2_0-g^2\Lambda^2
\label{equ0}
\end{equation}
Where \(m^2_0\) is the bare mass, and
\(\Lambda\) is the cut-off scale. Because both bare mass and the cut-off
scale are around \(10^{18} GeV\), in order to have a small weak scale 
renormalized mass \(m^2\), we need a fine-tuning mechanism.

Similar cases are widely existed in renormalizations and various
mixing mechanisms. The naturalness principle requires
that, any realistic model won't need  too much fine-tuning, it also
requires that the lagrangian parameters can not choose the values that will result in
excessive fine-tuning. It is one of the main reasons that 
we prefer the supersymmetric standard model, and it is also the main consideration
when we build a neutrino mass mechanism. These tasks requires a naturalness criteria that 
can reflect the level of naturalness correctly.  

The sensitivity criteria proposed by R. Barbirei and G.F.Giudice et al.\cite{BG}
is the first widely adopted quantitative indicator of the naturalness level.
Its idea is quite simple, If \(x_0\) is a input lagrangian parameter at the grand unification scale,
and \(y\) is a computed observable output parameter like masses, Yukawa couplings etc at the weak scale,
if we varies lagrangian parameter \(x_0\) at the grand unification scale, 
the corresponding computed weak scale observable parameter \(y\) will be varied, 
Barbieri and Giudice's sensitivity criteria \(c\) is defined as:

\begin{equation}
c=\vert \frac{\Delta y/y}{\Delta x_0/x_0}\vert=\vert \frac{\partial \ln y}{\partial \ln x_0}\vert
\label{equ2}
\end{equation} 

Here need to emphasis that we usually need to consider the effect of a parameter on
another parameter that has different canonical dimension. we prefer a dimensionless 
\(c\), thus Barbieri and Giudice chooses  \(\Delta y/y\) as the basis of comparison. 

Barbieri and Giudice set \(c\approx10\) as the naturalness cut-off, 
any \(c\) much greater than \(10\) will be classified as fine-tuned. Since then
hundreds authors apply this criteria to various problems, from setting a 
naturalness contour for SUSY particle search\cite{Azuelos:2002qw},
to the fine-tuning problem of the neutrino
seesaw mechanism\cite{Casas:2004gh}.
Barbieri and Giudice's sensitivity criteria has been widely adopted 
as the doctrine of naturalness judgment. 

But Barbieri and Giudice's sensitivity criteria is not reliable, many examples\cite{GWA}\cite{CS}
show that it fails for certain cases, because the sensitivity criteria plays 
an important role in new model building, it is worth to investigate the relationship
between the naturalness and the sensitivity, and find a correct and reliable criteria.

In mathematics, naturalness can be classified as a type of initial condition sensitivity
problem. Find how large the probability is for a output parameter in a
certain range of values is the best way to describe the initial condition sensitivity.
If we have a system with an input parameter \(x_0\) and a corresponding output 
parameter \(y\), If the probability of the input parameter around the value \(x_0\) 
is much smaller than the probability of the output parameter around the value \(y\),
then the system is initial condition sensitive, which means need fine-tuning in physics.

If we assume the input parameter \(x_0\) has a uniform probability distribution, 
then the output parameter \(y\) will have a probability distribution of 
\(\vert \partial x_0/\partial y\vert \),
if we choose \(\vert \partial y/\partial x_0\vert \), the inverse function of the output parameter 
probability distribution 
as a criteria to judge fine-tuning, then  those parameter regions that will result in
too small probability distribution \(\vert \partial x_0/\partial y\vert \), or too big 
its inverse function \(\vert \partial y/\partial x_0\vert \) can be classified as the 
fine-tuned region. 

Because we need to consider the fine-tuning property with two different types 
of parameters, for example, compare masses with the gauge couplings, and we also want 
a dimensionless fine-tuning indicator that does not depend on the parameters we are comparing. 
So Barbieri and Giudice's sensitivity criteria chooses the logarithmic function 
\(\vert \partial \ln y/\partial \ln x_0\vert \) as a fine-tuning criterion.  
Although \(c\) becomes dimensionless, while changing from \(\vert \partial y/\partial x_0\vert \)
to \(\vert \partial \ln y/\partial \ln x_0\vert \) means we have changed the assumption that
the probability distribution of the input parameter \(x\) from the uniform distribution to 
\(y(x_0)/x_0\) distribution, which means we assumed that lagrangian parameters tend to choose
certain values at the grand unification scale. Obviously this will greatly influenced the 
naturalness judgment.

For example, if we apply Barbieri and Giudice's 
formula (Eq.~(\ref{equ2})) to the \(\Phi^4\) model (Eq.~(\ref{equ-1})).
Clearly from Eq.~(\ref{equ0}) we know the scalar mass is highly fine-tuned. 
But on the other hand, if we integrate the mass renormalization group equation, we have:

\begin{equation}
m^2=m^2_0 \exp (\int_0^t(g^2/16\pi^2-1)dt)
\label{Prob2}
\end{equation}

Apply Barbieri and Giudice's sensitivity definition to Eq.~(\ref{Prob2}), it gives a result of 
sensitivity \(\partial\ln m^2/\partial\ln m^2_0\) equals to one, and it is not fine-tuned.

Obviously, the origin of this problem is whatever energy scale it is, the ratio \(\Delta m^2/m^2\)
is always fixed, Chosen the logarithumic function of the parameter rather than the parameter 
itself as a basis of comparison means we assume the variation ratio of the langrangian 
parameter  rather than the parameter itself is even probability distributed at the grand unification 
scale.   Thus even \(\Delta m^2\) is only around hundreds \(GeV^2s\) while \(\Delta m^2\) is 
around  \(10^{36} GeV^2\), we still think they are equivalent. Generally, 
the consequence of this is, if any parameter runs as an 
Exponent function \(y=y_0\exp(f)\) as energy scale \(t\) changes, if the exponent 
\(f=f(t)\) happens to be a function  of the energy scale \(t\) only, then even \(f\) is  
very large, and \(y\) blows up so quickly as \(t\) increased 
Barbieri and Giudice's sensitivity criteria will still give a sensitivity equals to \(1\) result, 
which tells you that the pure scalar field is not fine-tuned. Only when the index \(f\) is 
not only the function of \(t\), but also the function of \(y_0\),
then will Barbieri and Giuice's sensitivity criteria give a not equal to one sensitivity.

Return to the renormalization, normally the exponent \(f\) is consists of a constant part, 
several anonymous dimension parts \(\gamma_i\), some of the anonymous dimension parts are
depend on the lanrangian parameters \(x_0\), some are not,  although those parts of exponent 
that not depended on \(x_0\) will contribute the fine-tuning,  
it will still be ignored by Barbieri and Giudice's sensitivity criteria.
The anonymous dimension part reflects the relative changing of various parameters, and will depend on 
the initial conditions like masses, coupling constants,
it is obvious, if we adopted Barbieri and Giudice's criteria to calculate 
the sensitivity, this part will give a not equal to one result. 
so in the fact, What Barbieri and Giudice et al.'s sensitivity criteria judged is not the initial
condition sensitivity, but the anonymous dimension sensitivity, 
which reflects how much a parameter depends on the relative changes of  the various parameters.
So we'd better call Barbieri and Giudice et al.'s sensitivity ``anonymous sensitivity'' rather
than the initial condition sensitivity.

Besides the above mentioned problem, the sensitivity criteria have many other problems.
After Barbieri and Giudice's naturalness criteria has been proposed, G. Anderson et al\cite{GWA} 
first pointed out that under certain circumstances Barbieri and Giudice's 
naturalness criteria failed to give a correct result that consistent with known phenomena. 
P. Ciafaloni et al\cite{CS} also gave examples show that Barbieri and Giudice's naturalness
judgment is not valid under certain circumstances. 

The example given by  G. Anderson et al\cite{GWA} is regarding the high sensitivity
of \(\Lambda_{QCD}\) to the strong coupling constant \(g\),

\begin{equation}
\Lambda_{QCD}=M_P\exp{(-\frac{(4\pi)^2}{bg^2(M_P)})}
\label{equ8}
\end{equation}

Apply Barbieri and Giudice et al's definition of naturalness indicator we 
can calculate the sensitivity of \(\Lambda_{QCD}\) to the strong coupling 
constant \(g\) at the grand unification scale:
\begin{equation}
C(g)=\frac{4\pi}{b}\frac{1}{\alpha_s(M_P)}
\label{equ9}
\end{equation}
This value is greater than 100, much larger than the naturalness upper
bound set by Barbieri and Giudice. but actually it 
is protected by gauge symmetry, and is not fine tuned.

Carefully examed examples similar to the large sensitivity of \(\Lambda_{QCD}\),
we found all these examples occur when comparing parameters with different canonical
dimensions. Mathematically, comparing two parameters with different canonical dimensions  
is difficult, Barbieri and Giudice did aware this difficulty, and introduced the 
logarithmic function to rescale each parameter to a dimensionless formation, they thought this could be 
sufficient to eliminated the effects of the scale difference and dimensional difference,
but these examples show that, this method can not cancel these effects.

We know generally, there's a Gaussian fixed point at the origin of the parameter space
for renormalization group equations. around the origin, 
If we rescale the momenta by a factor of \(\Lambda\), then two different parameters 
\(\tau\) and \(h\) can be expanded as: 
\begin{equation}
\tau\approx\Lambda^{\alpha}\tau_0
\end{equation}
\begin{equation}
h\approx\Lambda^{\beta}h_0
\end{equation}
Here \(\alpha\) and \(\beta\) are corresponding canonical dimensions.
Because of the renormalization, the scale \(\Lambda\) will link these two parameters together.
even they may not have any other relations.
If we calculate the sensitivity of \(h\) to the variation of 
\(\tau\), approximately, it would be:

\begin{equation}
\frac{\partial h}{\partial \tau}\approx -\frac{\beta}{\alpha}\frac{h}{\tau}
\end{equation}

This effect was known as scaling effect in statistical physics, which exists anywhere 
when two parameters have different canonical dimensions, obviously,
it has nothing to do with the fine-tuning. but when
we convert both parameters to dimensionless parameters by Barbieri and 
Giudice's technique, the factor of different canonical dimension \(-\beta/\alpha\)
is still there, this is because the scaling phenomena is not linear, it can not be eliminated by 
Barbieri and Giudice's technique. 

If one parameter \(\tau\) has a marginal canonical dimension, then we can not use the 
above argument, we must consider the 
higher order term. take various couplings renormalization as an example:
\begin{equation}
\tau\approx(1+\alpha\tau_0\ln\frac{1}{\Lambda})\tau_0
\end{equation}

Similarly, the result would be:
\begin{equation}
\frac{\partial  h}{\partial \tau}\approx\frac{\beta}{\alpha\tau}\frac{h}{\tau}
\end{equation}

\begin{figure}
\includegraphics[angle=90, width=0.5\textwidth]{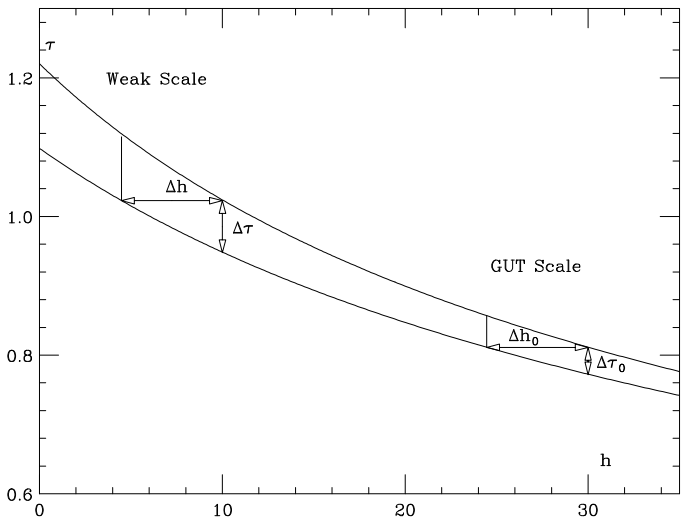}
\caption{RG trajectories in phase space spanned by two parameters with different dimension}
\label{Fig1}
\end{figure}

We can define a dimensional effect factor 
\(\Delta=\beta h/\alpha\tau^2\) or
 \(\beta h/\alpha\tau^2\) for later reference. In the parameter space if two parameters have 
the same dimension then \(\Delta\) becomes one, otherwise this factor may become significant,
A schematic diagram is shown in Fig.\ref{Fig1}.

If we look the naturalness problem from the phase diagram spanned by all the parameters 
(Fig.\ref{Fig1}),
There is a small area in this phase diagram which represents the weak scale, and 
there is also an area represents the grand unification scale, the renormalization flows 
run from the grand unification scale area and go to the weak scale area,
the ``naturalness'' requires that, whatever initial condition we choose, the weak scale area
is always smaller or, maybe a little large than the grand unification scale area, 
even the ratio \(\Delta h_0/\Delta\tau_0\) is big.
So it is better to understand the naturalness principle as the weak scale stability rather 
than the small sensitivity. High sensitivity doesn't mean unnatural.

The effect of different canonical dimensions is widely existed, for example, at low temperature,
the Plank radiation law becomes 
\(E_{\nu}=\frac{8\pi h\nu^3}{c^3}e^{-h\nu/kT}\),
If we calculate the sensitivity of \(E_{\nu}\) to the variation of the temperature, consider
\(T\approx 4k\), and \(\nu\approx 10^{15}Hz\), then you
will have an extremely large sensitivity \(c\approx 10^4\), but we never doubt the correctness 
of the Plank radiation law.

Besides these problems, The sensitivity criteria also implied  the relationship between 
the input and the output is monotonic, only the input parameter \(x\) can lead to the output
parameter \(y\), there's no such circumstance that both input parameters \(x_1\) and \(x_2\) 
will eventually lead to  a same output parameter \(y\).

Although the relationships between parameters linked by most renormalization group equations
are monotonic, while most mixing cases are not, one output parameter is usually
corresponding to two input parameters. For example, mixing of \(M_z\) mass and \(M_W\) mass, 
mixing of CP-even Higgs masses in Supersymmetric Standard Models, and mixing of fermionic masses etc. 

Take the mixing of \(M_z\) mass in MSSM model as an example, calculate the \(M_z\) mass 
at the initial condition (at grand unification scale) \(m=200, M=40, \tan\beta=18 \), 
and gradually reduced mass \(m\), then we will find the non-monotonic relationship between 
grand unification scale variable \(m\) and weak scale variable \(M_z\).

Obviously, it is not a monotonic function of \(m\), for example, if  weak scale mass
\(M_z\) is around \(80GeV\), there are two grand unification scale parameter regions that
can contribute this result, one is around \(m\) is \(80GeV\), the other region is around
\(m\) is \(150GeV\). Barbieri and Giudice's definition only counted one region's contribution
, it will overestimate the naturalness level.
We should count all possible GUT scale parameters contributions.

A good definition of naturalness criteria should be able to solve all the problems
listed above. Because the problems listed above, we can not choose the logarithmic function, 
instead, we need to use \(\partial y/\partial x_0\)
directly.  refers to the definition of Lyapunov exponent, which used to define the
initial condition sensitivity in dynamical systems, we can write down the \(t\) evolution of the
probability distribution to the variation of \(x_0\):

\begin{equation}
\frac{\delta y}{\delta x_0}=\Delta_0 e^{\lambda t}
\end{equation}

The dimensionless factor \(\lambda\) reflects the shrinkage of the probability, 
here \(\Delta_0\) is a kind of background probability density at the grand unification scale 
which need to be subtracted,
mathematically,
\(\lambda\) reflects the level of \(y\) fine-tuning when \(x_0\) changes. 
when the parameters \(x\) and \(y\) have the same 
canonical dimension, \(\Delta_0\) becomes one,  also considering the 
non-monotonic propriety, finally we can define a Lyapunov exponent like index \(\lambda\) 
for the naturalness criteria:
\begin{equation}
\lambda=\frac{1}{t}\ln\sum\vert\frac{1}{\Delta_0}\frac{\delta y}{\delta x_0}\vert
\end{equation}
If it is not monotonic, we divided and sum over all monotonic regions.

Although the fine-tuning criteria problem in high energy physics is somewhat
similar to the problem of using Lyapunov exponent to judge whether a 
nonlinear system is chaos or not, these two situations also have  
important difference. 
In nonlinear physics, if the Lyapunov exponent is negative, then the phase
space shrinks while time increases, it is obviously not initial condition sensitive,
thus it is not chaos, if the Lyapunov exponent is greater than zero, then the system is
initial condition sensitive and will be classified as chaos.
Similarly, when we consider the fine-tuning problem in high energy physics, 
if \(\lambda<0\), for the same reason, we can easily classify the system as not
initial condition sensitive of not fine-tuning, but for the cases that have
\(\lambda>0\), the situation is a little more complex, this is because for systems 
in nonlinear physics, 
the time variable \(t\) can go to infinity while in high energy physics,
the running parameter \(t\) can not go beyond the grand unification scale, 
which is around \(38\). so for the situations that have
small positive \(\lambda\), even the range of grand unification parameter space 
is a little bigger than the weak scale parameter space, it still can be 
thought as not initial condition sensitive, or not fine-tuned. So we should define
a reasonable positive fine-tuning upper limit for \(\lambda\).

In probability theory people usually define the probability \(p<0.05\) as 
small probability event and can be considered as hard to happen, although this is a more strict 
condition than Barbieri and Giudice's
sensitivity less than \(10\) criteria, we still adopt this doctrine, and define a upper limit
for the fine-tuning. Suppose fine-tuning occurs when \(p<0.05\), that means \(\exp(\lambda t)=1/0.05\),
we immediately have the upper limit of the \(\lambda\) index is 0.08.
According to this definition, all parameters with \(\lambda <0.08\)  will be safe
and not fine-tuned, and if \(\lambda > 0.08\) we learn that it is less than 5\% of chance
to have this weak scale value thus quite impossible. Not like Barbieri and Giudice's
naturalness \(c\approx 10\) cut-off, which doesn't have any physical meaning, our method
gives a clear physical meaning of the naturalness cut-off.

Numerical calculations with both Barbieri and Giudice's sensitivity criteria and \(\lambda\)
show that, for the cases with zero engineering dimension \(4-\frac{3}{2}n_f-n_b\) and both 
parameters have the same canonical dimension, and the relations are monotonic,  
the difference is not significant, this is because the effect of anonymous dimensions \(\gamma_i\)
are similar.  for MSSM model, when \(M=40 GeV, m=83.5GeV\) and \(\tan\beta =18\), sensitivity 
of \(m_h\) to the variation of \(m\) equals to 1, while new criteria when 
\(m=68.5GeV\) \(\lambda\) becomes 
positive.  But for \(\Phi^4\) model scalar mass,
which has engineering dimension equals to \(1\), sensitivity \(c=1<10\), while 
\(\lambda=1+\frac{g^2_{GUT}}{32\pi^2}\), greater than \(0.08\).
For the large sensitivity of \(\Lambda_{QCD}\), it is not difficult to calculate that
\(\lambda=(\ln\frac{\Lambda_{QCD}}{gM_P})/t\), which is far less than \(0.08\).
for non-monotonic case \(M_z\), if \(M_z=89.05GeV\), which corresponding to \(m=188.5GeV\) and
\(m=65.1GeV\) at the grand unification scale,  we calculated that sensitivity 
\(c=0.499\) and \(c=0.910\) respectively. while \(\lambda\) for \(M_z=89.05GeV\) is \(-0.046\).

In this paper we have investigated the widely adopted sensitivity criteria of naturalness, 
we found when comparing parameters with different canonical dimensions the sensitivity
usually will be very big, this should be understood as the scaling effect rather than the 
fine-tuning, under these circumstances the sensitivity is larger than the level of 
naturalness fine-tuning. When comparing parameters with the same canonical dimensions 
what we get is a type of ``anonymous sensitivity'', the result may be larger
or may be smaller than the true level of naturalness fine-tuning. 
All the calculations based on sensitivity
criteria become unreliable. In summary, the widely adopted 
sensitivity criteria is not reliable, can not truly reflect the naturalness properties.
We defined a new criteria to solve all the problems the sensitivity criteria has, and also
gives a clear physical meaning to the naturalness cut-off value.

\acknowledgements{The author would like to thank Dr. G.W. Anderson for advising, 
This work was supported in part by the US Department of Energy,
Division of High Energy Physics under grant No. DE-FG02-91-ER4086.}

\end{document}